\documentclass{article}

\usepackage{microtype}
\usepackage{graphicx}
\usepackage{subfigure}
\usepackage{booktabs} % for professional tables
\usepackage{color}

\newcommand{\cotwo}{\textrm{CO}_2}

\usepackage{hyperref}

\usepackage[accepted]{icml2024}

\usepackage{amsmath}
\usepackage{amssymb}
\usepackage{mathtools}
\usepackage{amsthm}
\usepackage{listings}
\usepackage[capitalize,noabbrev]{cleveref}

\theoremstyle{plain}

\theoremstyle{definition}

\theoremstyle{remark}

\usepackage[textsize=tiny]{todonotes}

\icmltitlerunning{Fourier Neural Operator based surrogates for $\cotwo$ storage in realistic geologies}

\begin{document}

\twocolumn[
\icmltitle{Fourier Neural Operator based surrogates for $\cotwo$ storage in realistic geologies}




\begin{icmlauthorlist}
\icmlauthor{Anirban Chandra}{shell_siti}
\icmlauthor{Marius Koch}{nvidia}
\icmlauthor{Suraj Pawar}{shell_siti}
\icmlauthor{Aniruddha Panda}{shell_sim}
\icmlauthor{Kamyar Azizzadenesheli}{nvidia}
\icmlauthor{Jeroen Snippe}{shell_sgsi}
\icmlauthor{Faruk O. Alpak}{shell_siep}
\icmlauthor{Farah Hariri}{nvidia}
\icmlauthor{Clement  Etienam}{nvidia}
\icmlauthor{Pandu Devarakota}{shell_siti}
\icmlauthor{Anima Anandkumar}{caltech,nvidia}
\icmlauthor{Detlef Hohl}{shell_siti}

\end{icmlauthorlist}

% Shell affiliations
\icmlaffiliation{shell_sgsi}{Shell Global Solutions International B.V., Amsterdam, Netherlands}
\icmlaffiliation{shell_siti}{Shell Information Technology International Inc., Houston, TX, USA }
\icmlaffiliation{shell_siep}{Shell International Exploration and Production Inc., Houston, TX, USA}
\icmlaffiliation{shell_sim}{Shell India Markets Pvt. Ltd., Bangalore, India }

% Caltech affiliations
\icmlaffiliation{nvidia}{NVIDIA, Santa Clara, CA, USA.}

% NVIDIA affiliations
\icmlaffiliation{caltech}{Department of Computing and Mathematical Sciences, California Institute of Technology, Pasadena, CA, USA}

\icmlcorrespondingauthor{Anirban Chandra}{anirban.chandra@shell.com}
\icmlcorrespondingauthor{Pandu Devarakota}{pandu.devarakota@shell.com}

\icmlkeywords{Machine Learning, ICML}

\vskip 0.3in
]



\printAffiliationsAndNotice{}  % leave blank if no need to mention equal contribution

\begin{abstract}
This study aims to develop surrogate models for accelerating decision making processes associated with carbon capture and storage (CCS) technologies.  Selection of sub-surface $\cotwo$ storage sites often necessitates expensive and involved simulations of $\cotwo$ flow fields. Here, we develop a Fourier Neural Operator (FNO) based model for real-time, high-resolution simulation of $\cotwo$ plume migration. The model is trained on a comprehensive dataset generated from realistic subsurface parameters and offers $\mathcal{O}(10^5)$ computational acceleration with minimal sacrifice in prediction accuracy. We also explore super-resolution experiments to improve the computational cost of training the FNO based models. Additionally, we present various strategies for improving the reliability of predictions from the model, which is crucial while assessing actual geological sites. This novel framework, based on NVIDIA’s Modulus library, will allow rapid screening of sites for CCS. The discussed workflows and strategies can be applied to other energy solutions like geothermal reservoir modeling and hydrogen storage. Our work scales scientific machine learning models to realistic 3D systems that are more consistent with real-life subsurface aquifers/reservoirs, paving the way for next-generation digital twins for subsurface CCS applications.
\end{abstract}

\section{Introduction}
Carbon dioxide ($\cotwo$) capture and storage (CCS) is a critical technology for decarbonization and achieving net zero emissions and climate goals. The first step of CCS involves the separation of $\cotwo$ from the gases produced by large power stations or industrial plants. It is then transported to locations where it can be stored in geological formations below the surface. The process involves integration of $\cotwo$ capture, compression, transport, and storage in the subsurface. The theoretical capacity of $\cotwo$ storing in deep geological formations globally is vast and far exceeds that required to reach net-zero emissions \cite{rogelj2018mitigation, de2014carbon, kelemen2019overview}. Currently only a small fraction of the potential global storage capacity is  used. A major challenge in identifying the most suitable sites for  storage is proving the containment of $\cotwo$  within a reservoir post-injection. This is because $\cotwo$  can migrate over time and potentially escape from the reservoir, posing a risk to the environment. In addition, the reservoir pressure propagation caused by $\cotwo$  injection needs to be carefully managed to avoid surface seismicity hazards. We need technological advancements and breakthroughs to rapidly screen potential sites for $\cotwo$  storage in a cost-effective and timely manner. 

Historically, numerical simulations have been used to model fluid flow and other physical phenomena in subsurface reservoirs and aquifers \cite{aziz1979petroleum, ertekin2001basic}. Existing reservoir simulators have been extended to model $\cotwo$  storage scenarios, but they can be computationally expensive, as they require solving non-linearly coupled partial differential equations (PDEs) \cite{nghiem2004modeling, wei2009estimate}. This can limit their use for studying the behavior of individual sites, especially when the subsurface properties are uncertain. The challenge of computational efficiency becomes even more pronounced for CCS site screening applications. This is because a substantial number of simulations are typically required to quantify the effects of subsurface uncertainties. Uncertain subsurface variables can include the dip of the formation, rock and fluid properties, and a wide range of reservoir engineering parameters that govern the trapping of $\cotwo$ in the subsurface. To accurately quantify the uncertainty surrounding the migration of the plume and the buildup and dissipation of pressure, it is necessary to carry out a large number of simulations.

Several surrogate models have been developed in the recent past for accurately capturing the dynamics for $\cotwo$ plume migration in the subsurface \cite{wen2021ccsnet, witte2022industry,grady2023model,falola2023rapid,witte2023fast}. In this study, we build on successes of FNO-based models~\citep{azizzadenesheli2024neural} and extend it to larger and more practical subsurface representations using NVIDIA's Modulus package. We train our models on an extensive dataset of numerical simulations performed using a realistic set of subsurface parameters. Along with  surrogate models that are $\mathcal{O}(10^5)$ faster than numerical simulations, we present and evaluate several physics based accuracy metrics that are relevant for assessing and monitoring CCS sites, such as $\cotwo$ plume migration and total mass in the reservoir. In an attempt to improve the computational efficiency of our models, we explore super-resolution, a niche of operator methods \cite{li2020neural,li2020multipole,kovachki2023neural}. Additionally, we propose several strategies to enhance the reliability of the models' predictions, which is vital when evaluating actual geological sites for $\cotwo$ storage such as outlier detection and enforcing mass conservation using physics informed loss functions ~\citep{li2021physics}. 

This paper is structured as follows: In Section \ref{sec:prob_setup} we discuss the problem setup along with details of the numerical flow simulations results that are used as training/testing data. Section \ref{sec:methods} describes our machine learning methodology. We present and discuss  results in Section \ref{sec:results}.

\section{Problem setup and dataset creation \label{sec:prob_setup}}

\begin{figure}[ht]
\vskip 0.2in
\begin{center}
\centerline{\includegraphics[width=0.8\columnwidth]{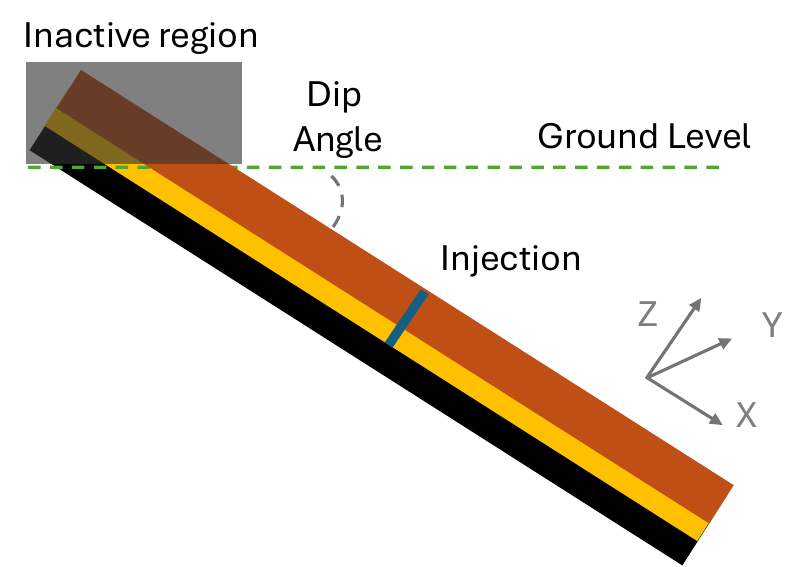}}
\caption{Schematic of the layered reservoir geometries. Different colors correspond to varying permeability. The parts of the reservoir protruding above ground level (reservoir ceiling), shaded in black, represent void-blocks (inactive regions), and changes based on reservoir dip angle and other input parameters (first row of Table \ref{tab:params}). }
\label{fig:schematic}
\end{center}
\vskip -0.2in
\end{figure}

To emulate realistic geological scenarios, we consider a three-dimensional dipping box model with a layered stratigraphic architecture. A schematic of the setup is shown in Figure \ref{fig:schematic}. For simplicity, a 2D representation of the box is used in the schematic visualization; $X,~Y,$ and $Z$ dimensions are $100$km, $3$km, $0.3$km respectively. The set of nine input parameters considered in this study are reported in Table~\ref{tab:params}, which are selected based on advice from domain experts. Outputs are $\cotwo$ mass accumulation ($m_{\cotwo}$), gas saturation ($S_g$), and change in pressure ($\delta p$). Two-phase Darcy flow equations govern the dynamics of fluid flow within the reservoir \cite{aziz1979petroleum}. 

After a grid sensitivity study we arrive at a final resolution that has $241 \times 10 \times 200$ grid points in the $X,~Y,$ and $Z$ direction, respectively. The chosen grid is the lowest resolution that captures the plume characteristics with the accuracy necessary for site assessment. While creating different samples (training dataset), the active region of the geometry changes based on the input parameters. In certain cases, the $\cotwo$  plume reaches the edge of the active region and accumulates because of the aquifer boundary condition employed in the simulation.

The number of grid points is constant across all samples leading to a simple data ingestion pipeline.  We train the surrogate model to predict the solution at eight time snapshots (including the end-of-injection time) in the post-injection period. These time snapshots are temporally spread out in a logarithmic manner until the end-of-simulation time, which is assumed to be a sufficiently long time at which migration of the $\cotwo$ plume approximately comes to a halt. Our spatial grid is also non-uniformly spaced with finer resolution near the injection location. For modeling of $\cotwo$ storage, we use $20$ years for injection period and $1000$ years post-injection. Our ML models are trained to predict $\cotwo$ plume migration after injection period. Such long time forecasting has not been carried out in previous machine-learning based $\cotwo$ plume migration studies that have predominantly focused on the injection period and forecasting up to only $\mathcal{O}(10)$ years \cite{wen2021ccsnet,wen2022u,wen2023real,jiang2023fourier}.

\begin{table}[t]
\caption{Scalar parameters used for parameterizing $\cotwo$ storage model and surrogate model development.}
\label{tab:params}
\begin{tabular}{p{5.3cm}p{2.1cm}}
\toprule 
\textbf{Parameters} & \textbf{Type} \vspace{0.05cm} \\  \toprule 
Permeability, number of geological layers, heterogeneity scale, reservoir height & Geology (layering)       \\ \hline
Reservoir dip angle & Geology         \\ \hline
Porosity & Geology \\ \hline
Vertical-horizontal permeability ratio & Geology                  \\ \hline
Reservoir Pressure & Geology, Fluid  \\ \hline
Reservoir Temperature & Fluid                    \\ \hline
\end{tabular}
\end{table}

Using a proprietary reservoir simulator for modeling multiphase multicomponent flow in porous media \cite{alpak2018variable, por1989fractured, regtien1995interactive} we generate $10,330$ samples for training and $2,593$ for testing. A separate validation set is not considered in this study as model performance was fairly consistent across the chosen hyperparameter combinations. Our choice of hyperparameters was based on previous studies of similar systems in simplified geometries \cite{pawar2023efficient, wen2021ccsnet}. Combinations of the nine input parameters are chosen using Latin hypercube sampling. Ranges of these parameters are documented in the Appendix \ref{SI:input_vars} (Table \ref{tab:paramsSI}). Output variables of interest are: (a) $\cotwo$ gas saturation - $S_g$ (b) $\cotwo$ mass accumulation - $m_{\cotwo}$ (c) Change in pressure from the initial state - $\delta p$.

\section{Machine learning methods \label{sec:methods}}
\subsection{Neural operators}
Neural operators are capable of learning the relationships between function spaces. They can map any parametric functional input to its corresponding output function. Operator learning here involves predicting one function based on other functions. In a physical system characterized by partial differential equations (PDEs), the input functions could be the initial condition $u_0$ defined on the physical domain $x$, the boundary condition $u_b$ and the forcing term $f$, defined on physical temporal domain $(x,t)$. The output in this case would be a PDE solution $u(x,t)$ at all $x$'s and $t$'s, where $x$ and $t$ are the space and time coordinates, respectively. For our problem, we define the domain $D \in R^d$ be a bounded and open set; $V$  be the input function space defined on $D$ that takes values in $R^{d_v}$; and $U$ be the output function space on $D$ that takes values in $R^{d_u}$. The mapping between the input function $v$ and output function $u$ is denoted by an operator
\begin{equation}
\mathcal{G}:\mathcal{V} \ni v \rightarrow u \in \mathcal{U}.
\end{equation}

We aim to approximate $\mathcal{G}$ with an operator $\mathcal{G}_{\theta}$ from training data 
$\mathcal{T}~=~\{(v^{(1)},u^{(1)}),(v^{(1)},u^{(2)}),\dots (v^{(m)},u^{(m)})\}$, where $v^{(i)}$ is drawn from a probability measure $\mu \in \mathcal{V}$ and $u^{(i)}~=~G(v^{(i)})$. Here $\theta$ refers to the trainable parameters of the neural operator and we learn this operator $\mathcal{G}_{\theta}$ by minimizing the following problem with a cost functional $\mathcal{C}$ 
\begin{equation}
\textrm{arg}\min_\theta \mathbb{E}_{v \in \mu} \big[\mathcal{C} \big(\mathcal{G}_{\theta}(v), \mathcal{G}(v) \big) \big].
\end{equation}

The Fourier neural operator (FNO) is an iterative architecture that employs kernel integral operations to learn mapping between two function spaces \cite{li2020fourier}. In FNO, the kernel integral operator is substituted with a convolution operation defined in the Fourier space, i.e., the coefficients of the Fourier series of the output functions are learned from the data. The FNO consists of a lifting layer, a point wise operator that maps the input co-dimension to higher dimensional representation using a fully connected neural network, then several Fourier layers, and finally the projection layer that maps the high-dimensional output co-dimension of the last Fourier layer to the output function. Several modifications have been proposed in the literature to apply FNOs to input and output functions defined on various domains and to handle complex geometry \cite{lu2022comprehensive}.  

For the output of the $l$'th Fourier layer $z_l$ with $d_v$ channels, the Fourier layer can be written as follows:
\begin{equation}
\mathcal{F}^{-1} \big( \mathcal{R}_l \cdot \mathcal{F} (z_l) \big)+W_lz_l
\end{equation}
where 
\begin{itemize}
\item $\mathcal{F}$ is the Fourier transform and is applied to each channel of $z_l$ separately. The higher modes of $\mathcal{F}(z_l)$ are truncated to retain only $k$ modes for computational efficiency.  Thus, $\mathcal{F}(z_l)$ has the shape $k \times d_v$.
\item The $W_l$ is yet another pointwise operator acting as a residual connection. 
\item We then multiply each mode index of $\mathcal{F}(z_l)$ with a learnable weight matrix (complex number) of shape $d_v \times d_v$ which form the weight matrix $\mathcal{R}_l\in \mathbb{C}^{(d_v \times d_v \times k)}$.
\item Finally, we perform inverse Fourier transforms and apply a point-wise nonlinear activation function of choice. When the discretization grids are regular, the approximate integral of Fourier transforms are carried out using fast Fourier Transform (FFT). In the inverse FFT$\mathcal{R}\cdot \mathcal{F}(z_l)$ zeros fill in the truncated modes.
\end{itemize}

The residual layer term is added to the output of each of the Fourier layers before it is passed through an activation function. This bias term allows FNOs to handle data with non-periodic boundary conditions as well as high frequency components. We use the relative $l_p$ loss to train the neural operator models. The loss function can be written as, 

\begin{equation}
L(u,\hat{u}) = \frac{\| (u - \hat{u}) \|_p}{\| u \|_p} + \alpha \frac{\|\int_V m_{\cotwo} - \int_V {\widehat{m}_{\cotwo}} \|_p}{\|\int_V  m_{\cotwo} \|_p},
\label{eqn:loss}
\end{equation}
where $u$ is the ground truth, $\hat{u}$ is the predicted output, $\int_V m_{\cotwo}$ is the ground truth for total $\cotwo$ mass, $\int_V \widehat{m}_{\cotwo}$ is the predicted total $\cotwo$ mass, $p$ is the order of norm, and $\alpha$ is the hyperparameter to assign a weight to the mass conservation penalty term.

\subsection{Software framework }
Our workflow is based on NVIDIA's open-source package Modulus \cite{NVIDIA_Modulus}. NVIDIA Modulus is a pytorch-based framework for scientific machine learning applications that  provides an extensive collection of network architectures and convenience functions for setting up training and inference pipelines. Several hardware and software optimizations are also implemented implicitly. In our work, we leverage the 4D FNO models that are available in Modulus' architectural library and heavily use \texttt{DistributedManager} for scaling to multi-GPU and multi-node settings. More details are provided in \ref{SI:modulus_architecture}.

\section{Results and Discussion \label{sec:results}}
\begin{figure*}[ht]
\vskip 0.2in
\begin{center}
\centerline{\includegraphics[width=0.8\textwidth]{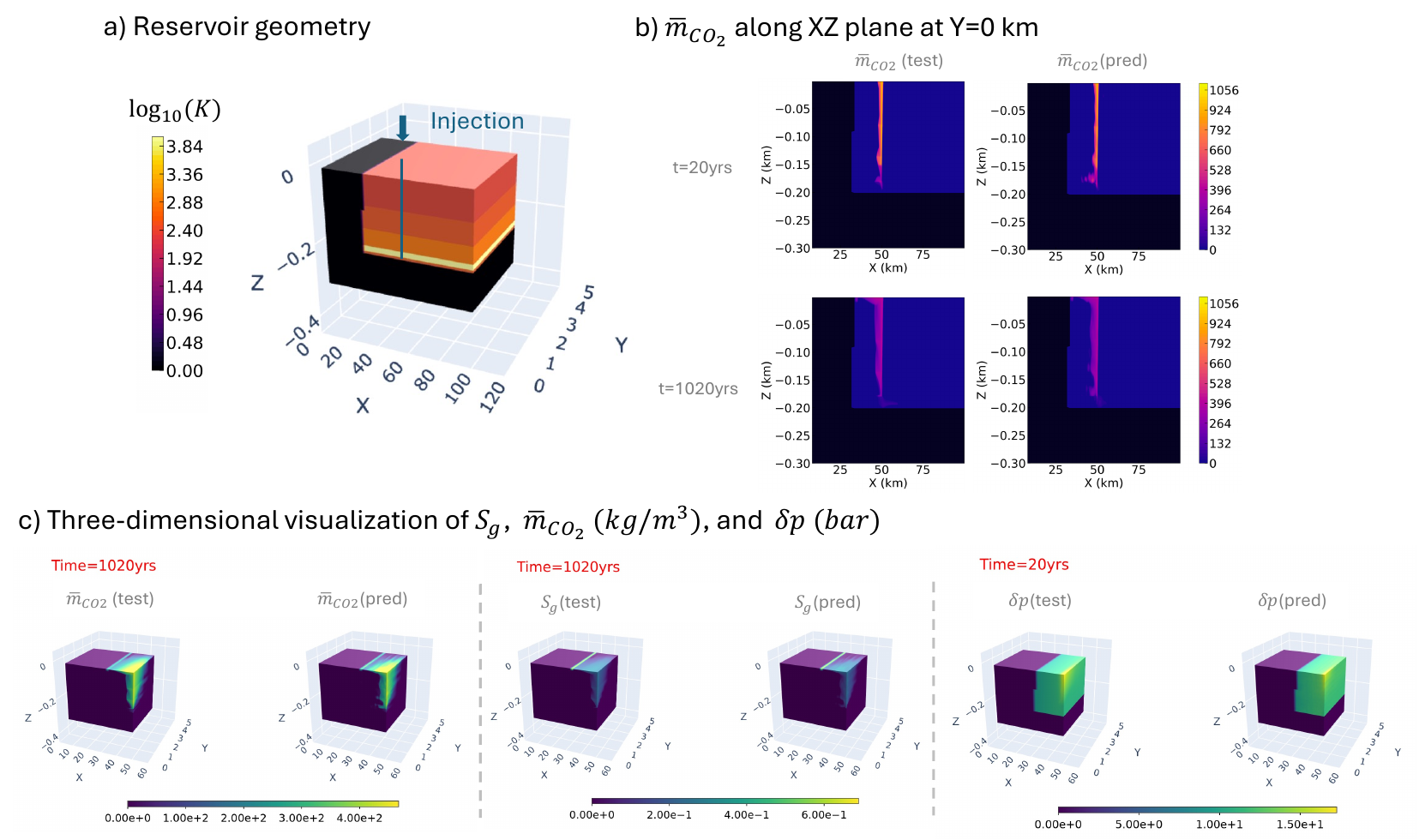}}
\caption{Visualization of fields. (a) 3D permeability map of the reservoir along with injection location. Distance is measured in km and permeability (K) in milliDarcy (mD). (b)  $\overline{m}_{\cotwo}$, measured in $kg/m^3$, as viewed on the XZ plane (c) Cross sectional views of various fields  at the injection location. $m_{\cotwo}$  and $S_g$ are shown at final time whereas $\delta p$ at initial time because of their respective importance in decision making.  
}
\label{fig:field_visualization}
\end{center}
\vskip -0.2in
\end{figure*}

Spatiotemporal distributions of mass of $\cotwo$ ($m_{\cotwo}$) and gas saturation ($S_g$) are two variables of interest, that are correlated to each other via phase densities. Another variable of interest is the pressure change compared to the initial pressure state,  $\delta p$.  Figure \ref{fig:field_visualization} shows 3D and 2D snapshots of a specific sample. For spatiotemporal visualization, we use $\overline{m}_{\cotwo}$, which is $m_{\cotwo}$ scaled by grid volume and porosity, to obtain a better visual representation of the plume. In all cases, a good match between ground truth (numerical simulation) and ML prediction is observed. We evaluate several global error metrics, both numerical and physical, to assess the accuracy of our predictions.

%%%%%%%%%%%%%%%%%%%%%%%
\subsection{Accuracy metrics}
\begin{figure*}[ht]
\vskip 0.2in
\begin{center}
\centerline{\includegraphics[width=0.8\textwidth]{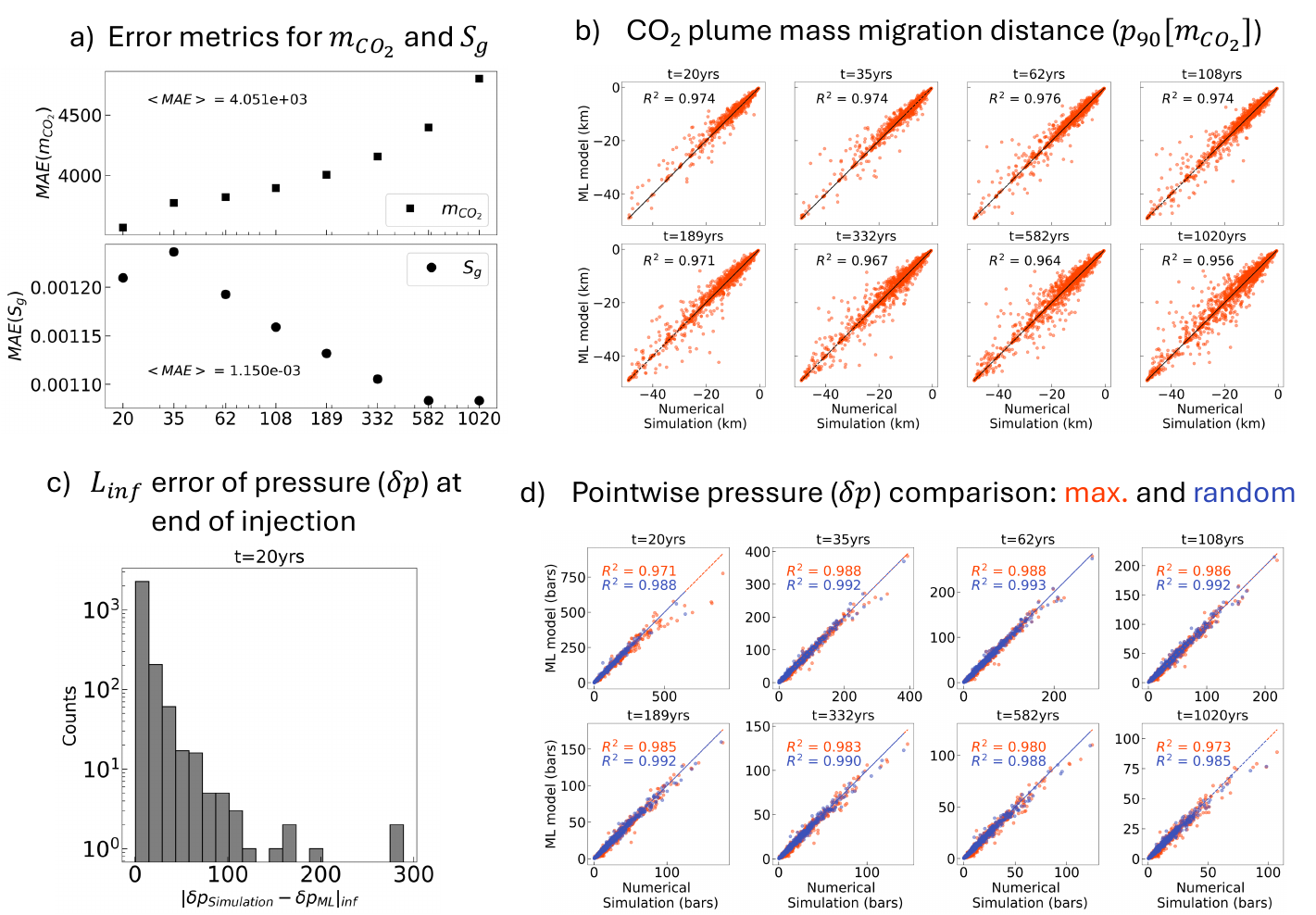}}
\caption{Error metrics (a) MAE for $m_{\cotwo}$  and $S_g$, with $<MAE>$ representing average in time (b) $R^2$ correlation plots of $90\%$ plume mass migration distance. (c) Maximum pointwise error in $\delta p$ (d) [red circles] $R^2$ correlation plots of $\delta p$ at maximum pressure location corresponding to the test sample. [blue circles] Correlation plots $\delta p$ for randomly selected locations within the active domain of the test sample. }
\label{fig:error_metrics}
\end{center}
\vskip -0.2in
\end{figure*}

Figure \ref{fig:error_metrics}(a) shows the mean absolute error (MAE) for the two solution variables of interest; MAE is generally a good metric for assessing the accuracy of the variables over the entire domain. In addition, we monitor physical metrics such as, $p_{90}[m_{\cotwo}]$, which is the migration distance of the 90\% plume mass contour from the injection location. In Figure \ref{fig:error_metrics} (b) we show the $R^2$ correlation plots of this metric at different points in time. With increasing time, the accuracy decreases slightly, as the plume migrates. Metrics for  pressure ($\delta p$) need more careful consideration as global values are not informative. Instead, we explore a few point-wise (local) metrics for assessing the accuracy of the pressure predictions of the FNO based model. Histograms of maximum point-wise error in $\delta p$ across samples are shown in Figure \ref{fig:error_metrics} (c). While most cases have a low error, there exists a $\mathcal{O}(10)$ outliers. Although this metric provides us with an upper bound, it is not a metric that cannot be easily evaluated in realistic settings. While monitoring CCS sites, we are often interested in the location where the maximum pressure occurs -- typically near the injection well. Thus, we evaluate the prediction from the model at  the location where the true pressure is maximum. Red circles in Figure \ref{fig:error_metrics}(d) demonstrate the accuracy of the FNO model with respect to this metric. Furthermore, for monitoring of CCS sites, the pressure buildup at additional (site-specific) locations can be important. To emulate this, while avoiding  a location bias, we randomly select a few additional locations from our test set and assess predictions from our model at the same location, as represented by the blue circles in Figure \ref{fig:error_metrics}(d). In both metrics, we observe a $R^2$ score greater than 0.97, suggesting that the model would provide reasonable predictions in most cases.

%%%%%%%%%%%%%%%%%%%%%%%
\subsection{Effect of weak imposition of mass conservation}
\begin{figure*}[ht]
\vskip 0.2in
\begin{center}
\centerline{\includegraphics[width=0.65\textwidth]{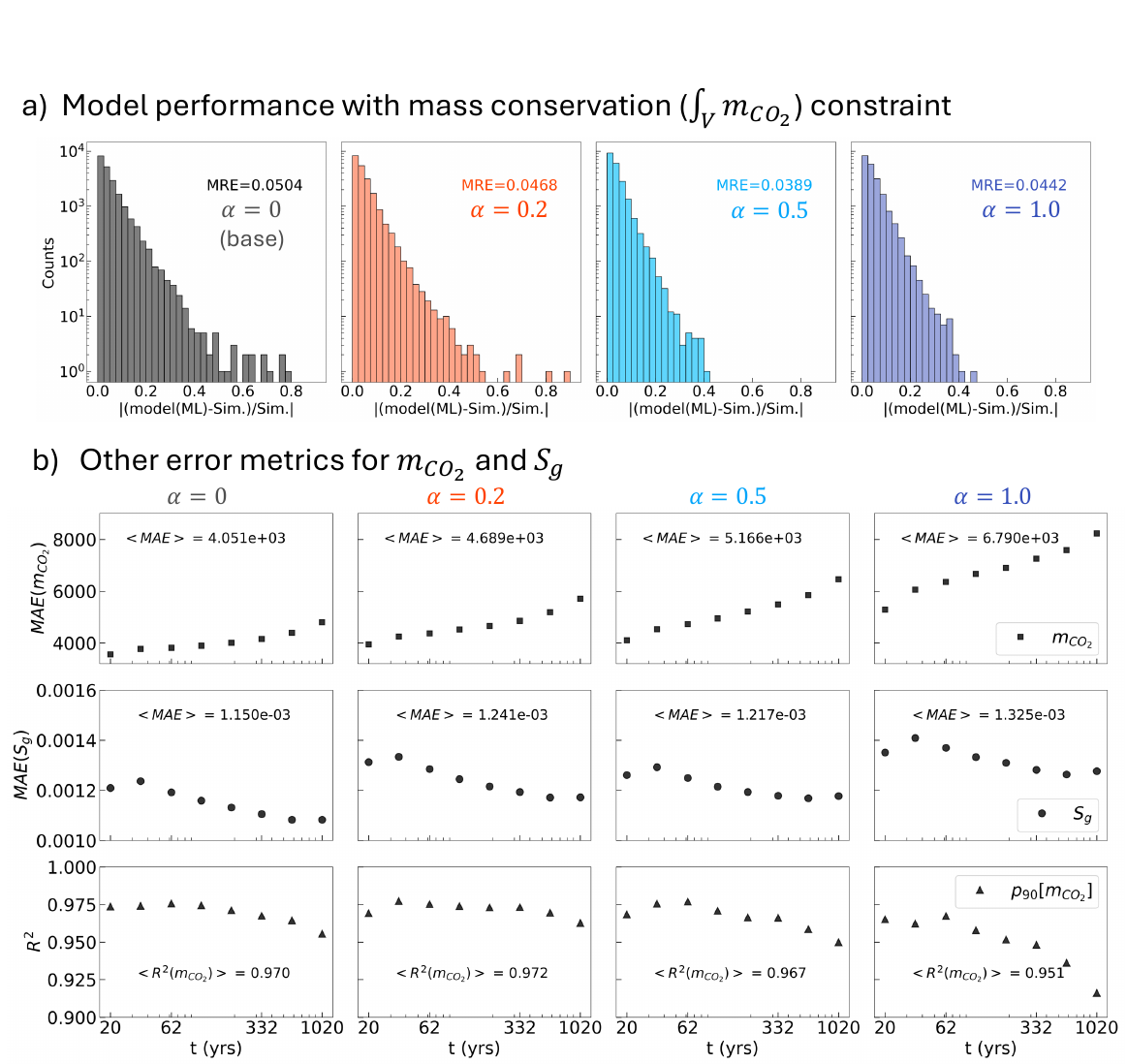}}
\caption{Performance metrics when a soft constraint on total mass in the system is imposed (a) Mean relative error (MRE) of total mass, $\int_V m_{\cotwo}$,  over all time instances as weighing factor in the loss function ($\alpha$) is varied. (b) MAE for $m_{\cotwo}$ and $S_g$ along with $R^2$  of $p_{90}[m_{\cotwo}]$}
\label{fig:mass_consv}
\end{center}
\vskip -0.2in
\end{figure*}

While our models exhibit good accuracy in all the metrics discussed in the previous subsection, an additional important metric to assess model fidelity is $\cotwo$ mass conservation, $\int_V m_{\cotwo}$.  As we generally train our models with the relative $L_2$ loss as shown in Equation \ref{eqn:loss} (with $\alpha$=0), there is no explicit constraint on the total mass conservation. Therefore, to enforce better mass conservation, we explore loss functions with finite values of $\alpha$. Figure \ref{fig:mass_consv}(a) shows the distribution of relative errors in mass conservation for four  settings of $\alpha$. The histograms are constructed considering all time instances. In the base case, i.e. $\alpha=0$, not only is the mean relative error (MRE) the highest, but several outliers exist. Here, MRE is used instead of MAE in this case to provide a more intuitive measure of the mismatch in mass conservation. Amongst all the considered values of $\alpha$, MRE reaches a minimum $\alpha = 0.5$. Other metrics are shown in Figure \ref{fig:mass_consv}(b). $MAE(m_{\cotwo})$ increases with increasing $\alpha$ which is intuitive as the additional soft constraint indirectly modifies this metric. At $\alpha=0.5$, an empirical optimum is achieved, where $MAE(S_g)$ and $p_{90}[m_{\cotwo}]$ is comparable to the base case. At the highest value of $\alpha$,  these error metrics become worse, which is indicative of a harder optimization problem. Other engineered loss functions, akin to those explored by \cite{wen2022u}, could further improve accuracy of predictions.

\subsection{Super resolution}
A key difference between neural operators and neural networks is the ability to learn functional mappings rather than finite-dimensional approximations. Thus, a model trained on a coarse-resolution data can yield predictions at a finer resolution - this is regarded as super resolution. For investigating the accuracy of super-resolution experiments, the ML models are trained and tested on a variety of resolutions. $ML(r_t,r_i)$ denotes that the model was trained on resolution $r_t$ and inferred on $r_i$, where $r_t$ and $r_i$ are selected from the set of resolutions $\{c_v, c_d, f \}$. $c_v$ and $c_d$ are volume-averaged and discrete representations of the variables of interest, respectively. Both types of down-sampling reduces the number of grid points by a factor of two in the y-direction. $f$ is the original simulated (ground truth) data on the finer grid. For comparison, the original simulated data, $Sim(f)$, is also down-sampled in the aforementioned ways, and is represented by $Sim(c_v)$ and $Sim(c_d)$. It should be noted that only the variables of interest (input and outputs) undergo two types down-sampling operations. The spatial grid is always discretely downsampled. More information regarding downsampling is provided in Appendix \ref{SI_downsampling}.

\begin{figure*}[ht]
\vskip 0.2in
\begin{center}
\centerline{\includegraphics[width=0.9\textwidth]{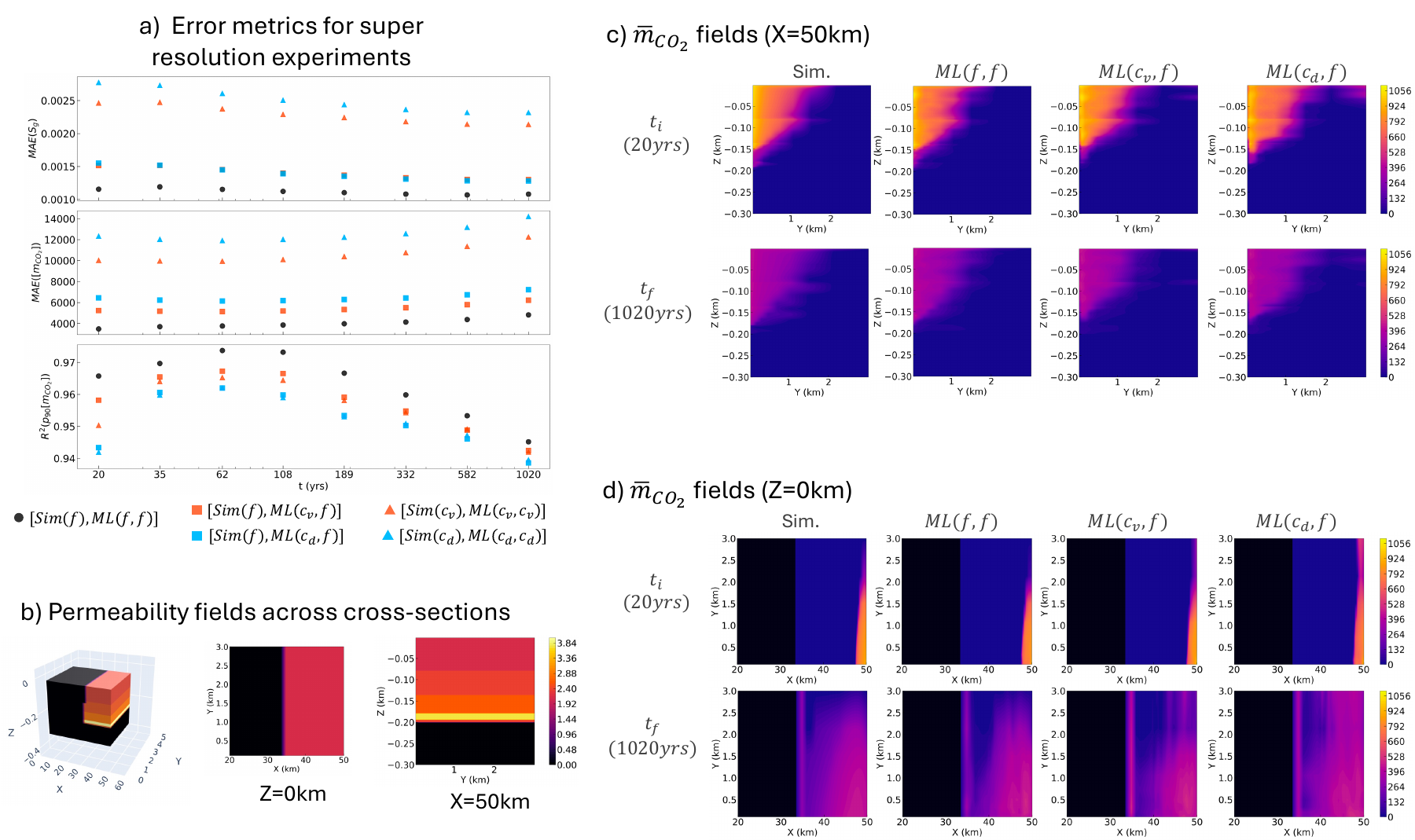}}
\caption{Super resolution experiments (a)   Error metrics for different cases – base case (black circles), model trained on vol. avg coarse grid and evaluated on fine grid (orange square), model trained on vol. avg coarse grid and evaluated on vol. avg coarse grid (orange triangle), model trained on discretely down-sampled coarse grid and evaluated on fine grid (blue square), model trained on discretely down-sampled coarse grid and evaluated on discretely down-sampled coarse grid (blue triangle),  (b, c, d) 2D visualization of fields along the Y axis.
}
\label{fig:superres}
\end{center}
\vskip -0.2in
\end{figure*}

Figure \ref{fig:superres}(a) shows the performance of models trained and evaluated at various resolutions. All the metrics -  MAE for the $MAE(S_g)$, $MAE(m_{\cotwo})$, and $R^2(p_{90}[m_{\cotwo}])$ - are evaluated with respect to numerical simulations. $[Sim(f), ML(f,f)]$, represented by black circles, is the base case where the model was trained and evaluated on the finer grid, and compared with simulated data on the finer grid. Similarly, $[Sim(f), ML(c_v,f)]$, denoted by orange squares, signify that the model was trained on a volume-averaged coarse grid and inference was performed on a fine grid, and compared against finer-grid numerical simulations to obtain the different error metrics. In all cases and through all time-instances, base case performs the best while $[Sim(c_d), ML(c_d,c_d)]$ performs the worst. The advantage of neural operators is predominantly demonstrated by the 'orange squares'. Even though these cases were trained on a volume-averaged representation of variables, the error metrics closely follow the base case. The $[Sim(f), ML(c_d,f)]$ comparison is slightly worse than the volume-averaged case as our grid is non-uniform and volume-averaging embeds more physical information.  Snapshots $m_{\cotwo}$ on two planes spanning the y-direction at the injection location  are shown in Figures \ref{fig:superres}(c) and (d). In all cases, $ML(c_v,f)$ represents $ML(f,f)$ more closely as compared to $ML(c_d,f)$. Errors are more prominent closer to the geometrical boundaries ($Y=30$km) due to coarser grid spacing.

\subsection{Outlier detection}
\begin{figure*}[!ht]
\vskip 0.2in
\begin{center}
\centerline{\includegraphics[width=1.\textwidth]{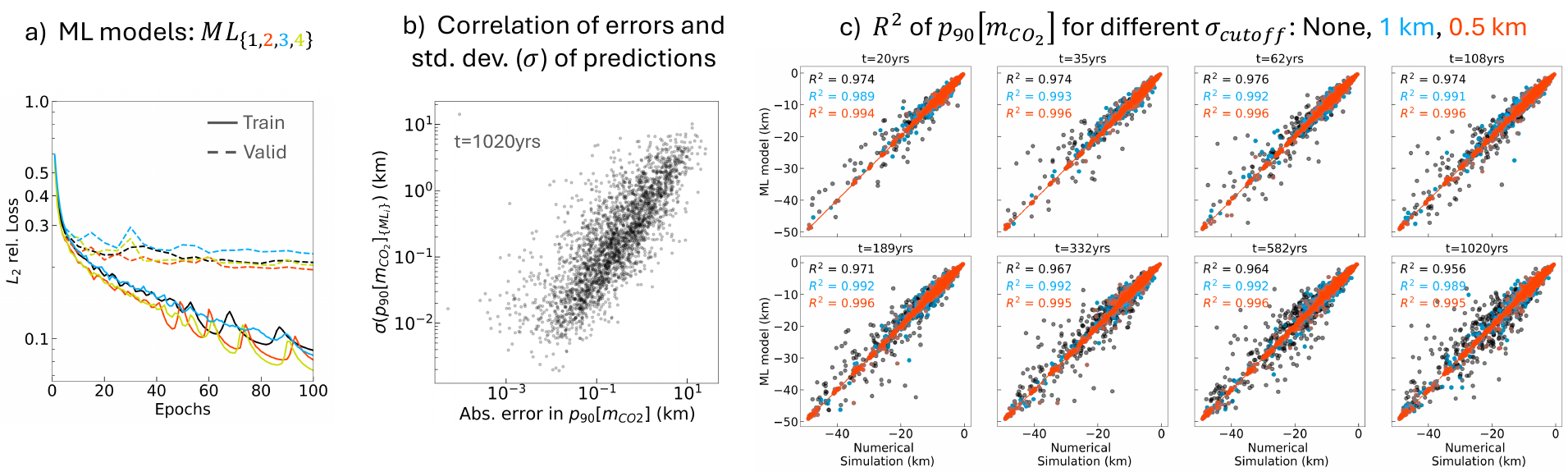}}
\caption{Detecting outliers based on ensemble models. (a)   Training statistics of four ML models with slightly varying hyperparameters. All four models have similar performance. (b) Variation of standard deviation of $90\%$ plume mass plume migration distance prediction from four ML models with absolute error in plume mass plume migration distance. (c) $R^2$ correlation of samples down-selected based on standard deviation of $90\%$ plume mass plume migration distance prediction}
\label{fig:outliers}
\end{center}
\vskip -0.2in
\end{figure*}

While the developed surrogate models allow rapid prediction of $p_{90}[m_{\cotwo}]$ for various geologic scenarios, in certain conditions, the predictions can be misleading. Generally, a slight reduction in overall accuracy is expected and acceptable due to geologic uncertainties, but highly deviating predictions can be misleading. In, Figure \ref{fig:outliers}, we present a methodology for detecting outliers based on model uncertainties. Slightly changing the hyperparameters (decoder width, Fourier modes, number of FNO layers) of the model leads to similar training and validation performance, as shown in Figure \ref{fig:outliers}(a). Details of the considered models are presented in Appendix \ref{SI:hyperparams}. The migration distance of the 90 percent of the (vertically integrated) $\cotwo$ plume mass, $p_{90}[m_{\cotwo}]$, is one of the most important metrics for CCS applications, thus we use this variable to tailor our outlier detection technique. In Figure \ref{fig:outliers}(b) we show that the standard deviation of the prediction of plume mass migration distance, $\sigma(p_{90}[m_{\cotwo}])$, has a strong positive correlation with the prediction error of this variable. While deploying the model (during inference), the error in $p_{90}[m_{\cotwo}]$ cannot be evaluated; therefore, this correlating behavior can be exploited to point out high error samples, i.e., the outliers. Figure \ref{fig:outliers}(c) shows $R^2$ correlations of plume mass migration distance of samples selected based on different $\sigma (p_{90}[m_{\cotwo}])$ values. No cutoff represents the base case; increasing value of $\sigma_{cutoff}$ represents a stricter removal of outlying samples and hence a reduction in the $R^2$ values. This methodology can be used to identify unreliable predictions from the surrogate model and run additional numerical simulations to enrich the training dataset as needed.  

\subsection{Model performance}
Lastly, we discuss the computational speedup of our model compared to numerical simulation. Training data  is generated  using an in-house proprietary simulation package for subsurface flows. The hardware used to produce a training sample consists of 4 physical cores of Intel Cascade Lake CPUs. The two ML models, $ML(m_{\cotwo}$,~$S_g)$ and  $ML(\delta p)$, are trained on 8 NVIDIA A100 GPUs and inferred on 1 NVIDIA A100 GPU.  Typically, our models are trained for 100 epochs. Table \ref{tab:performance} provides computational times for generating a sample of training data, training a model, and inference.  Inference times on a AMD EPYC 7763 CPU, which is $\mathcal{O}(20)$ slower than GPU inferencing, is presented as a reference;  in all our cases, we perform inference on GPU. The training and inference times for the ML model predicting $\delta p$ are similar to $ML(m_{\cotwo}$,~$S_g)$. Hyperparameters and architectural details are presented in  Appendix \ref{SI:hyperparams}. During GPU inference, using our ML model provides a speed-up  of $\mathcal{O}(10^5)$ with respect to numerical simulations.

\begin{table}[hbt]
\caption{Run times for training data generation, ML model ($S_g$, $m_{\cotwo}$) training and inference. Time is reported in either GPU or CPU (Intel Cascade Lake$^\ddagger$, AMD EPYC 7763$^\dagger$) seconds. }
\label{tab:performance}
\begin{tabular}{p{4.2cm}p{3.2cm}}
\toprule 
\textbf{Case type} & \textbf{Run time} 
\vspace{0.05cm} \\  \toprule 
Simulation (per sample) & $3.037 \times 10^{4}$ CPU$^\ddagger$ sec       \\ \hline
ML infer. (per sample) & $1.780   \times 10^{-1}$  GPU sec   \\ \hline
ML infer. (per sample) & $3.329$     CPU$^\dagger$ sec   \\ \hline
ML train (per epoch) & $7.205 \times 10^{3}$   GPU sec \\ \hline
\end{tabular}
\end{table}

A fair and exact comparison between numerical simulation and ML models based on runtime is not possible, as they were run on different hardware. However, we  show a simple break-even analysis using  \textit{CPU hr} as computational unit for numerical simulation and  \textit{GPU hr} as unit for ML models. Total time for generating the entire training dataset accounts to $t_{gen} =3.037 \times 10^{4} \textrm{ CPU sec} \times 12\,923 \textrm{ samples} = 1.090 \times 10^5 \textrm{ CPU hr} $. The total times for training and validation of a single ML model are, 
$t_{train} =100 \textrm{ epochs}  \times 7.205 \times 10^{3} \textrm{ GPU sec} = 200.138 \textrm{ GPU hr}$ and $t_{valid} =   1.780 \times 10^{-1} \textrm{ GPU sec} \times 2593 \textrm{ samples} = 0.128 \textrm{ GPU hr} $ respectively. Therefore total time spent building the two ML models is, $t_{ML-develop}=t_{gen}  + 2(t_{train} + t_{valid})  = 1.094 \times 10^{5} \textrm{ hr}$. The factor of two is due to the fact that $ML(\delta p)$ is developed as a separate model, and both models require similar training and validation times. 

In a typical screening setting, around $20,000$ Monte Carlo simulations are required for generating a diverse range of scenarios. The total computational time for developing the ML models and inferencing $20,000$ samples sums up to, $t_{ML-develop} + (20,000 \times 2 \times 1.780 \times 10^{-1} \textrm{ sec}) = 1.094 \times 10^{5} + 1.978 \textrm{ hr} = 1.094 \times 10^{5} \textrm{hr}$.
Producing $20,000$ samples through numerical simulation requires $20,000 \times 3.037 \times 10^{4}  \textrm{ sec} = 1.687 \times 10^{5} \textrm{ hr}$. Therefore, just one site assessment justifies training an ML model. The fully-trained model can be applied to numerous screening tasks, amplifying the benefits of our approach.

\section{Summary and Conclusions}

In this study, we have developed a model based on the Fourier Neural Operators (FNO) for real-time, high-resolution simulation of $\cotwo$ plume migration. This model is trained on an extensive dataset derived from realistic subsurface parameters. During inference, we observe a speedup of $\mathcal{O}(10^5)$  over traditional numerical simulators of $\cotwo$ flow fields with minimal reduction in accuracy. Along with fast surrogate models, we present and assess several physics based accuracy metrics that are relevant for assessing and monitoring CCS sites. 
In an attempt to improve the computational efficiency of our models, we explore training on coarser grid, that is created by downsampling a simulation created on a fine grid, and finally evaluating on a finer grid. Our experiments suggest that training on a coarse grid and evaluating on a finer grid has better accuracy than a case where the model is trained and evaluated on a coarse grid. This is a niche capability of a neural operator. Using uniform grids would further improve the accuracy of the models. Additionally, we propose several strategies -- outlier detection, enforcing mass conservation -- to enhance the reliability of the model’s predictions, which is vital when evaluating actual geological sites. Our methodologies and strategies can be potentially extended with some domain-specific adaptations to other energy solutions such as geothermal reservoir modeling and hydrogen storage. Our work scales up scientific machine learning models to realistic 3D systems that more consistent with real-life subsurface aquifers/reservoirs, and builds a foundation for advanced screening tools for subsurface CCS applications.

\newpage \newpage

%\section*{Accessibility}
%Authors are kindly asked to make their submissions as accessible as possible for everyone including people with %disabilities and sensory or neurological differences.
%Tips of how to achieve this and what to pay attention to will be provided on the conference website \url{http://icml.cc/}.


\section*{Acknowledgements}
The authors thank Shell International Exploration and
Production Inc., Shell Information Technology International Inc., Shell India Markets Pvt. Ltd., and Shell Global Solutions International B.V.  for permission to publish this work.

%\textbf{Do not} include acknowledgements in the initial version of
%the paper submitted for blind review.

%If a paper is accepted, the final camera-ready version can (and
%usually should) include acknowledgements.  Such acknowledgements
%should be placed at the end of the section, in an unnumbered section
%that does not count towards the paper page limit. Typically, this will 
%include thanks to reviewers who gave useful comments, to colleagues 
%who contributed to the ideas, and to funding agencies and corporate 
%sponsors that provided financial support.

\section*{Impact Statement}
The goal of this work is to advance the field of AI for Science. The ethical impacts and expected societal implications are those that  are well established when advancing the field of Machine Learning, none which we feel must be specifically highlighted here.


\bibliography{bibliography}
\bibliographystyle{icml2024}

%%%%%%%%%%%%%%%%%%%%%%%%%%%%%%%%%%%%%%%%%%%%%%%%%%%%%%%%%%%%%%%%%%%%%%%%%%%%%%%
%%%%%%%%%%%%%%%%%%%%%%%%%%%%%%%%%%%%%%%%%%%%%%%%%%%%%%%%%%%%%%%%%%%%%%%%%%%%%%%
% APPENDIX
%%%%%%%%%%%%%%%%%%%%%%%%%%%%%%%%%%%%%%%%%%%%%%%%%%%%%%%%%%%%%%%%%%%%%%%%%%%%%%%
%%%%%%%%%%%%%%%%%%%%%%%%%%%%%%%%%%%%%%%%%%%%%%%%%%%%%%%%%%%%%%%%%%%%%%%%%%%%%%%
\newpage
\appendix
\onecolumn
\section{Appendix}

\subsection{Input variables \label{SI:input_vars}}

Ranges of various input variables are shown in Table \ref{tab:paramsSI}. All the reported values are scalars - permeability fields are generated using a proprietary algorithm scripted in our in-house flow simulator. 

\begin{table}[hbt]
\caption{Ranges of scalar parameters used for parameterizing $\cotwo$ storage model and surrogate model development}
\label{tab:paramsSI}
\begin{tabular}{p{0.7\textwidth}p{0.3\textwidth}}

\toprule 
\textbf{Parameters} & \textbf{Range} \vspace{0.05cm} \\  \toprule 
Base permeability ($\log_{10}$, base 1mD)
&    (1, 3.30) 
\\ \hline
Number of geological layers
&     (2, 20)$_d$   
\\ \hline
Heterogeneity scale
&     (0, 1)
\\ \hline
Reservoir height (m)
&  (15, 300)
\\ \hline
Reservoir dip angle (degrees)
&  (0, 6)
\\ \hline
Porosity 
&  (0.05, 0.35)
\\ \hline
Vertical-horizontal permeability ratio ($\log_{10}$)
&  (-6, 0)                 
\\ \hline
Reservoir Pressure (bar)
&  (100, 300)
\\ \hline
Reservoir Temperature $(^o C)$
&  (25, 135)
\\ \hline
\end{tabular}
\end{table}

\subsection{Architectural modifications \label{SI:modulus_architecture}}

In NVIDIA Modulus, FNO models can be constructed by a simple function call providing configurations such as the number of FNO layers, number of modes and activation functions. NVIDIA Modulus being open-source in combination with its modular design enables additional customisation of the FNO architectures. Listing~\ref{fno_cust} provides pseudo code for customising the lift layer akin to the models used in this work.

\begin{lstlisting}[language=Python, caption=Pseudocode for changing the lift layer of FNO to a linear transormation., label=fno_cust]
class CustomFNO4DEncoder(modulus.models.fno.FNO4DEncoder):
    def build_lift_network(self):
        self.lift_network = torch.nn.Sequential()
        self.lift_network.append(
            Linear(self.in_channels, self.fno_modes)
        )
\end{lstlisting}

Other methods used from NVIDIA Modulus are logging routines and the \texttt{DistributedManager}. The \texttt{DistributedManager}
provides a singleton class for setting up the parallel environment by assigning available devices,
carrying out optimisations like clearing device cache and automatically choosing the communication
backend based on the availability of NCCL.

\subsection{Model architecture and hyperparameters \label{SI:hyperparams}}
In all base cases domain size is nx, ny, nz, nt = 214, 10, 200, 8 and hyperparameters are shown in Table \ref{tab:hyperparams}.

\begin{table}[!hbt]
\centering
\begin{tabular}{|l|l|}
\hline
\textbf{Parameter} & \textbf{Value} \\
\hline
\multicolumn{2}{|c|}{\textbf{Decoder}} \\
\hline
layers & 1 \\
layer\_size & 128 \\
\hline
\multicolumn{2}{|c|}{\textbf{FNO}} \\
\hline
dimension & 4 \\
latent\_channels & 36 \\
fno\_layers & 6 \\
fno\_modes  & [10,5,10,5] \\
padding & 0 \\
\hline
\multicolumn{2}{|c|}{\textbf{Scheduler}} \\
\hline
initial\_lr & 1.E-3 \\
decay\_rate & .95 \\
decay\_epochs & 2 \\
\hline
\end{tabular}
\caption{Deep Learning Model Configuration}
\label{tab:hyperparams}
\end{table}

It should be noted that the number of input dimensions are 10 -- 6 field variables and 4 coordinates corresponding to x,y,z, and t. For super resolution experiments, dimensions are (nx, ny, nz, nt = 214, 5, 200, 8) and 2 modes are retained in the y direction. For outlier detection, the three additional models (in addition to the base case) are created by changing the following hyperparameters independently w.r.t the base case --  \texttt{padding=2, latent\_channels=48, fno\_layers=8}.

\newpage
\subsection{Visualization of various output variables}

\begin{figure*}[!ht]
\vskip 0.2in
\begin{center}
\centerline{\includegraphics[width=0.8\textwidth]{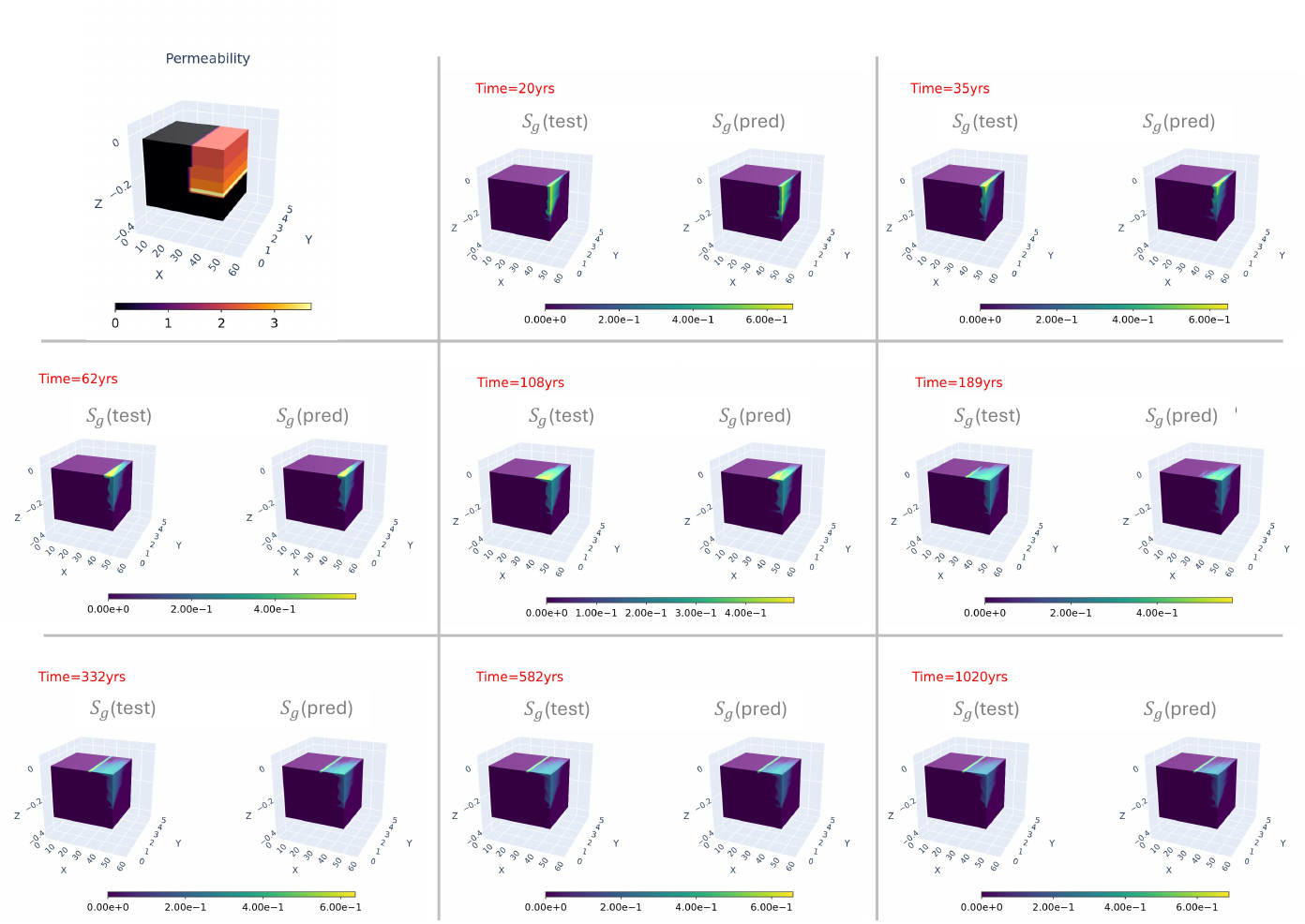}}
\caption{Comparison of $S_g$-test and $S_g$-pred of a specific sample at different time instances}
\label{fig-outliers}
\end{center}
\vskip -0.2in
\end{figure*}

\begin{figure*}[!ht]
\vskip 0.2in
\begin{center}
\centerline{\includegraphics[width=0.8\textwidth]{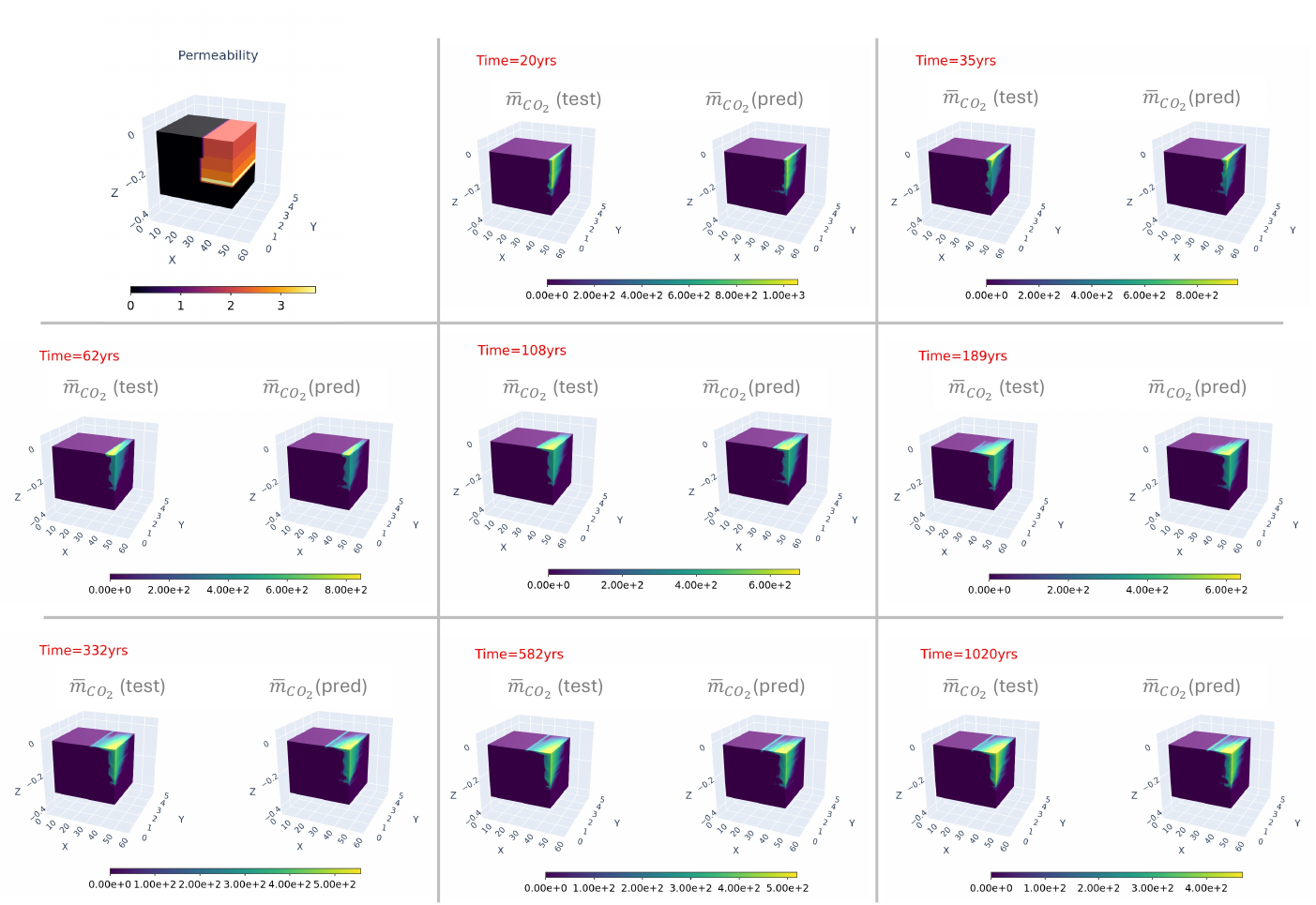}}
\caption{ Comparison of $\overline{m}_{\cotwo}$-test and $\overline{m}_{\cotwo}$-pred of a specific sample at different time instances}
\label{fig-outliers}
\end{center}
\vskip -0.2in
\end{figure*}

\begin{figure*}[!hbt]
\vskip 0.2in
\begin{center}
\centerline{\includegraphics[width=0.8\textwidth]{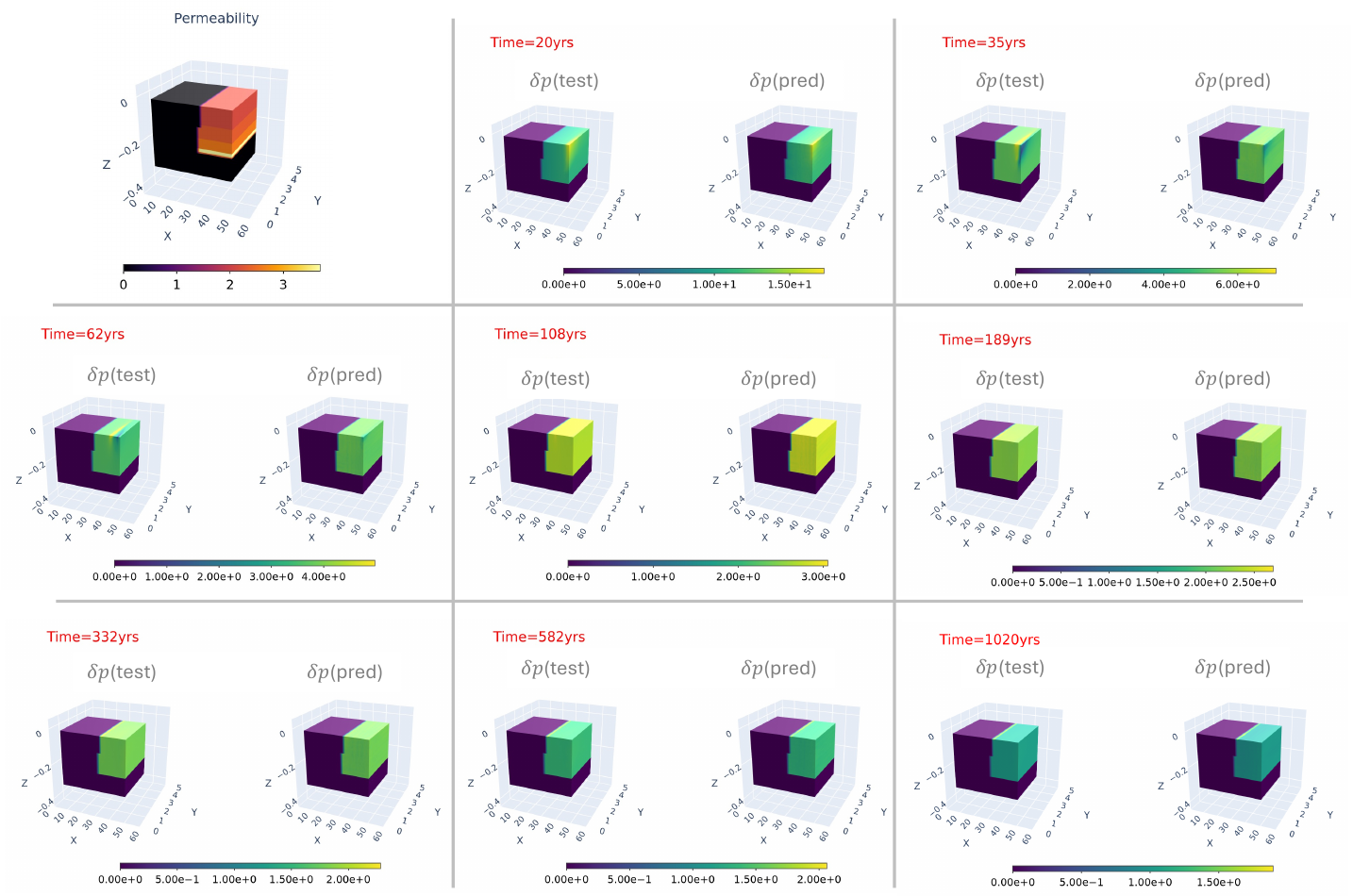}}
\caption{ Comparison of$\delta p$-test and $\delta p$-pred of a specific sample at different time instances}
\label{fig-outliers}
\end{center}
\vskip -0.2in
\end{figure*}

\newpage
\subsection{Mass Conservation}

\begin{figure*}[!hbt]
\vskip 0.2in
\begin{center}
\centerline{\includegraphics[width=0.8\textwidth]{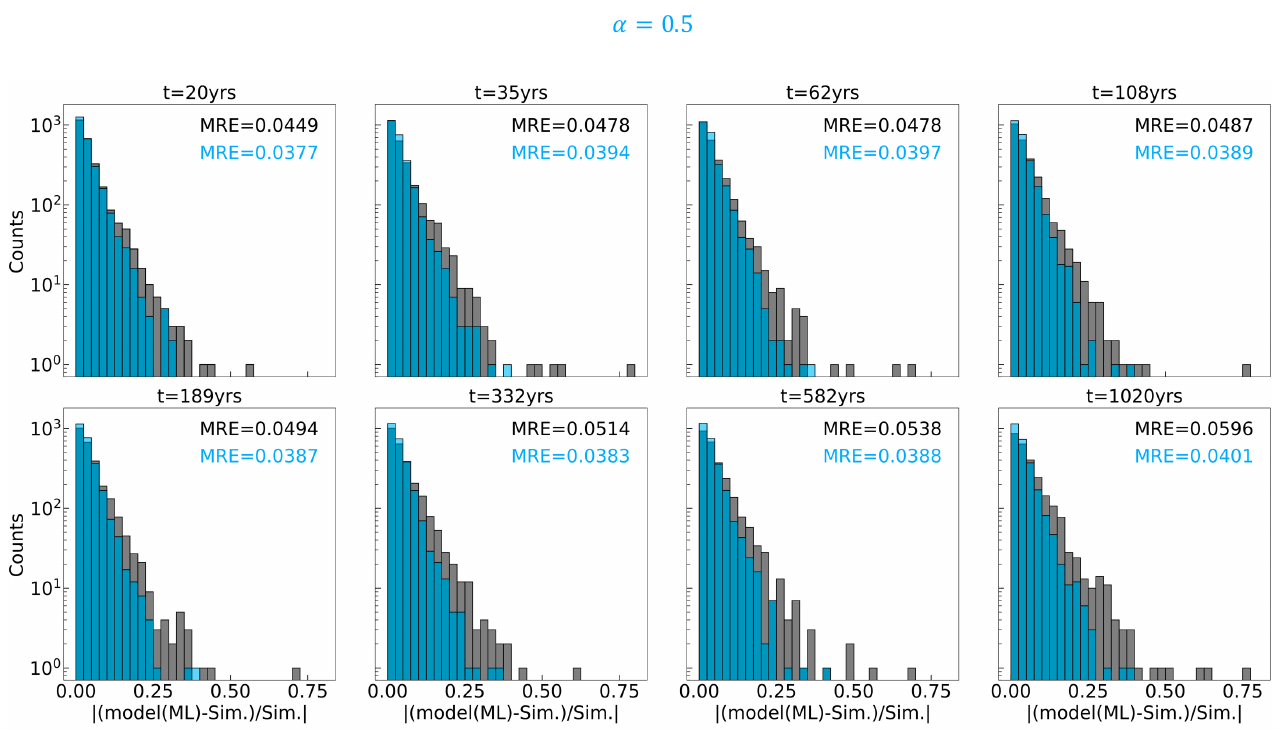}}
\caption{Mean relative error in total mass conservation at different times when $\alpha=0.5$}
\label{fig-outliers}
\end{center}
\vskip -0.2in
\end{figure*}

\subsection{Downsampling strategies \label{SI_downsampling}}
\begin{figure*}[h]
\vskip 0.2in
\begin{center}
\centerline{\includegraphics[width=0.3\textwidth]{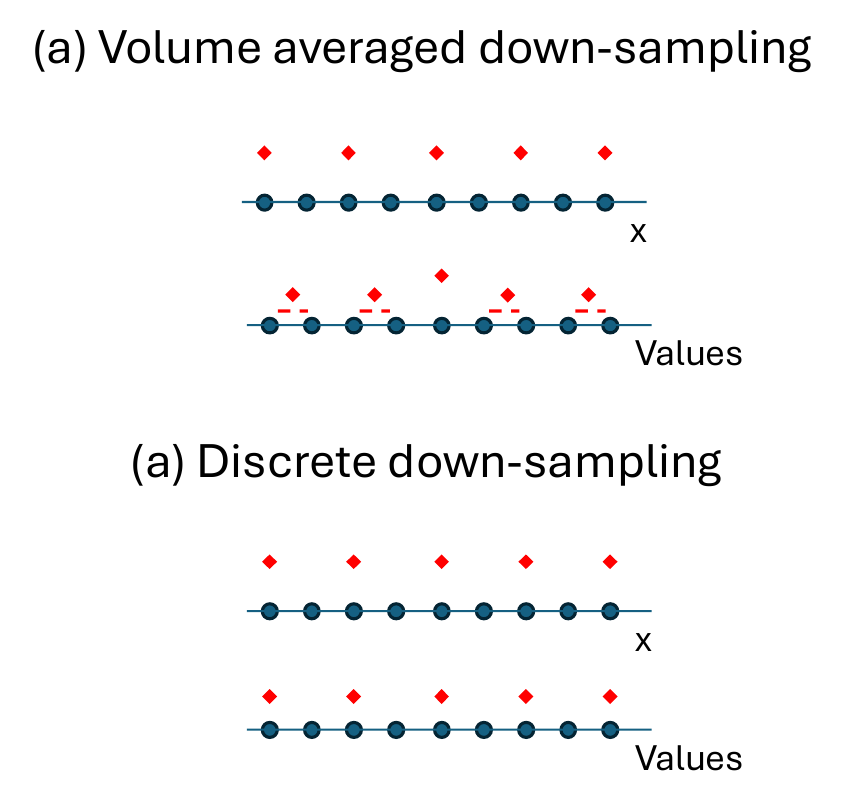}}
\caption{Different downsampling strategies}
\label{fig:SI_downsampling}
\end{center}
\vskip -0.2in
\end{figure*}

We choose to create our coarsened representation in two ways: (a) Volume averaged downsampling - The field variables are volume averaged (b) Discrete downsampling - every other point from the original grid is selected. A schematic of our methodology is shown in Figure \ref{fig:SI_downsampling}. Note that in all our cases, downsampling is only performed in the y-direction. The grid (X, Y, Z) coordinates is always discretely picked - no volume averaging is performed.

%%%%%%%%%%%%%%%%%%%%%%%%%%%%%%%%%%%%%%%%%%%%%%%%%%%%%%%%%%%%%%%%%%%%%%%%%%%%%%%
%%%%%%%%%%%%%%%%%%%%%%%%%%%%%%%%%%%%%%%%%%%%%%%%%%%%%%%%%%%%%%%%%%%%%%%%%%%%%%%

\end{document}